\documentclass[aps,rmp,twocolumn,groupedaddress,floatfix]{revtex4-1}
\pdfoutput=1
\usepackage{natbib}
\usepackage{amsmath}
\usepackage{amsfonts}
\usepackage{amssymb}
\usepackage{graphicx}

\newcommand{\eq}[1]{Eq.~(\ref{#1})}
\newcommand{\fig}[1]{Fig.~\ref{#1}}
\newcommand{\eqs}[2]{Eqs.~(\ref{#1}) and (\ref{#2})}
\newcommand{\figs}[2]{Figs.~\ref{#1} and \ref{#2}}
\newcommand{\app}[1]{Appendix~\ref{#1}}

\def \Ud{U_d}
\def \Un{U_n}
\def \Utot{U_\mathrm{tot}}
\def \Ns{Ns}
\def \NUd{N\Ud}
\def \NUn{N\Un}
\def \NUtot{N\Utot}
\def \Nsigma{N \sigma}
\def \hk{h_k}
\def \Udeff{\tilde{U}_d}
\def \Uneff{\tilde{U}_n}

\def \seff{\tilde{s}}
\def \Nseff{N\seff}
\def \NUdeff{N\Udeff}
\def \NUneff{N\Uneff}

\def \ftilde{\tilde{f}}
\def \wtilde{\tilde{w}}
\def \sprime{s'}
\def \Udprime{\Ud'}
\def \Unprime{\Un'}
\def \Nsprime{N\sprime}
\def \NUdprime{N\Udprime}
\def \NUnprime{N\Unprime}

\def \T2{\langle T_2 \rangle}
\def \Ne{N_e}
\def \Sn{S_n}

\begin{document}

\title{The equivalence between weak and strong purifying selection}

\author{Benjamin H. Good$^{1}$}
\author{Michael M. Desai$^{1}$}
\affiliation{\mbox{${}^1$Department of Organismic and Evolutionary Biology, Department of Physics, and} \mbox{FAS Center for Systems Biology, Harvard University}}

\begin{abstract}

Weak purifying selection, acting on many linked mutations, may play a major role in shaping patterns of molecular evolution in natural populations. Yet efforts to infer these effects from DNA sequence data are limited by our incomplete understanding of weak selection on local genomic scales.  Here, we demonstrate a natural symmetry between weak and strong selection, in which the effects of many weakly selected mutations on patterns of molecular evolution are equivalent to a smaller number of more strongly selected mutations. By introducing a coarse-grained ``effective selection coefficient,'' we derive an explicit mapping between weakly selected populations and their strongly selected counterparts, which allows us to make accurate and efficient predictions across the full range of selection strengths. This suggests that an effective selection coefficient and effective mutation rate --- not an effective population size --- is the most accurate summary of the effects of selection over locally linked regions. Moreover, this correspondence places fundamental limits on our ability to resolve the effects of weak selection from contemporary sequence data alone.

\end{abstract}

\date{\today}
\maketitle

Purifying selection maintains important biological function by purging deleterious mutations and is thought to play a major role in shaping the patterns of molecular evolution in many organisms \citep{charlesworth:2012}. In principle, these patterns can provide important information about the selective forces operating within a population, and could be used to disentangle this signal from other factors, such as demographic history \citep{williamson:etal:2005}. Yet existing methods are limited by our incomplete understanding of purifying selection on local genomic scales, where many linked sites are potentially selected against.

The action of selection on neighboring sites creates correlations within a genotype that can be difficult to disentangle from each other. Early treatments assumed that these correlations were essentially equivalent to an increase in genetic drift, or a reduction in effective population size, and that the individual sites otherwise evolve independently \citep{hill:robertson:1966,charlesworth:2009}. Recent studies of these ``Hill-Robertson interference'' effects have challenged the validity of this assumption \citep{santiago:cabellero:1998, bustamante:etal:2001, comeron:kreitman:2002, comeron:etal:2008}, particularly for the case of weak selection.  But without a simple alternative, the effective population size picture continues to dominate much of our qualitative understanding of linked selection and its application to data from natural populations.

Meanwhile, attempts to incorporate linkage more explicitly have been limited to the case where the strength of purifying selection is strong and the number of deleterious polymorphisms is small. In this regime, correlations within genotypes are still highly uncertain, but the distribution of fitnesses within the population can be modeled very precisely.  For extremely strong selection, this leads to the classic background selection picture, in which the apparent size of the population is reduced to the size of the least-loaded class \citep{charlesworth:etal:1993}. More generally, methods based on the structured coalescent framework \citep{kaplan:etal:1988} lead to improved (though more complicated) analytical predictions \citep{walczak:etal:2012,nicolaisen:desai:2012}, as well as a class of extremely efficient backward-time simulations that can be used to rapidly calculate any quantity of interest \citep{hudson:kaplan:1994,gordo:etal:2002}.

Yet there is increasing evidence that at these local genomic scales, selection is dominated not by a few strongly deleterious polymorphisms, but rather by many more weakly selected mutations that can segregate in the same lineage \citep{comeron:kreitman:2002, bartolome:charlesworth:2006, loewe:charlesworth:2007, barraclough:etal:2007, kaiser:charlesworth:2008, seger:etal:2010, lohmueller:etal:2011, subramanian:2012}. In this regime, the strong-selection results break down due to the increased importance of stochastic fluctuations, which can carry some deleterious alleles to intermediate or high frequencies while driving others to extinction. In the extreme case, these fluctuations can sometimes lead to the extinction of the wild-type class and the subsequent fixation of a deleterious allele --- an effect known as Muller's ratchet \citep{muller:1964}. The complexity of these forces has lead to the belief that the dynamics of weak selection are of a fundamentally different character than strong selection, and that a new theoretical picture is required to understand them. Various numerical methods have been devised for this regime, but they either become computationally prohibitive for more than a few selected sites \citep{krone:neuhauser:1997, neuhauser:krone:1997, barton:etheridge:2004, barton:etal:2004} or require computation time that scales with the size of the population \citep{ofallon:etal:2010, seger:etal:2010}, similar to traditional forward-time simulations.

The apparent intractability of the weak selection regime is somewhat paradoxical, given that the limit of \emph{infinitely} weak selection is simply a neutral population. Part of this difficulty arises from the fact that existing coalescent models of purifying selection explicitly track the number of deleterious mutations in each individual. An entirely neutral theory that required similar accounting for neutral mutations would be intractable for many of the same reasons. However, this superficial difficulty obscures a more fundamental aspect of weakly selected mutations: individually, they have a negligible impact on the ancestral process and are indistinguishable from their neutral counterparts, but the accumulation of many such mutations can have a significant effect on the overall diversity of the sample.

\begin{figure}[t]
\centering
\includegraphics[width=0.95\columnwidth]{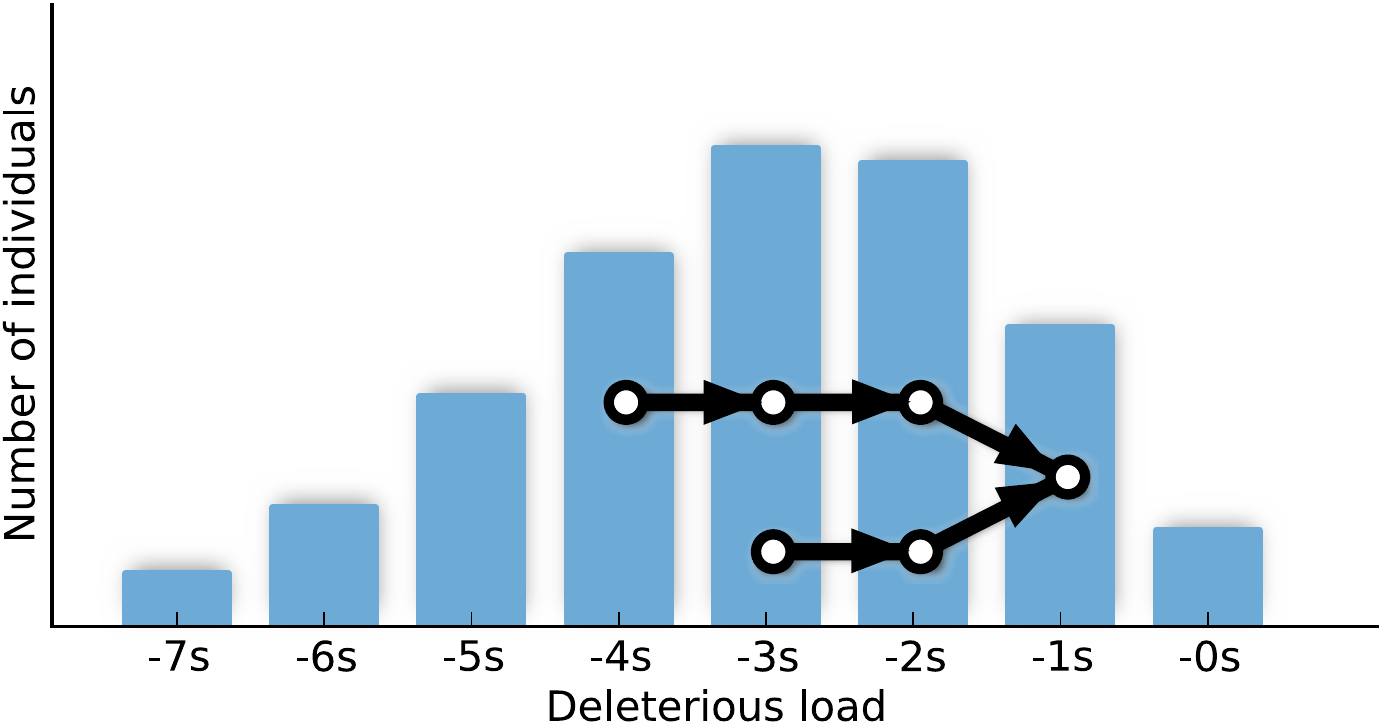}
\vspace{1em}
\caption{The deterministic prediction for the distribution of fitnesses in the population at mutation-selection balance when $\lambda = \Ud/s \approx 3$, and a possible ancestral history for a sample of two individuals. \label{fig:strong-selection-diagram}}
\end{figure}

In the present work, we exploit this separation of scales to establish a correspondence between the strong and weak selection regimes. By relaxing our definition of neutral and selected mutations and introducing a rescaled \emph{effective selection strength}, we demonstrate an equivalence principle relating the patterns of diversity among populations with differing strengths of selection.  For a given population in the weak selection regime, this defines a mapping to a corresponding strong-selection model that captures most of the quantitative features of the original population.  The previously developed strong selection results can therefore be extended to provide a single, unified theory valid over the entire range of selective effects, which provides valuable qualitative insights into the net effect of purifying selection.

This correspondence has obvious practical benefits for the analysis of DNA sequence data, since the existing strong-selection techniques can generate efficient predictions across a wide range of parameters, and can potentially form the basis for self-consistent inference of the underlying selective forces and population sizes. These results have important qualitative implications as well, providing a simple and intuitive alternative to the popular (yet flawed) effective population size picture. Rather, our correspondence suggests that a more natural local quantity is an \emph{effective strength of selection}, defined over some characteristic linkage block. However, the equivalence between strong and weak selection --- and the equivalence between weakly selected populations themselves --- suggests an inherent limit to our ability to resolve selection pressures from contemporary polymorphism data, especially at the level of individual sites.

\section{Analysis}

In order to quantify the molecular diversity generated by purifying selection at many linked sites, we confine our attention to a simple and well-studied model in which these effects are known to play a major role. We consider a population of $N$ non-recombining haploid individuals that accumulate neutral mutations at rate $\Un$ and suffer deleterious mutations with a constant multiplicative fitness effect $s$ at rate $\Ud$. We assume that the sequences in the population are well described by an infinite sites model in which each mutation occurs at a unique site in the genome, and we neglect compensatory or otherwise beneficial mutations. In addition, we work in the standard diffusion limit $N \to \infty$, where the scaled parameters $\Ns$, $\NUd$, and $\NUn$ are sufficient to determine all quantities of interest.

\subsection{Strong selection}
The behavior of this model has been well-characterized when selection against the deleterious alleles is sufficiently strong (we discuss the exact conditions below). In this case, the population reaches a steady state in which the continuous influx of deleterious mutations is balanced on average by the action of selection against them. In the limit $\Ns \to \infty$ where genetic drift can be neglected, the expected fraction of individuals with $k$ deleterious mutations (``fitness class $k$'') is given by \begin{equation}
\label{eq:fitness-distribution}
\hk =\frac{\lambda^k}{k!} e^{-\lambda} \, ,
\end{equation}
where $\lambda = \Ud/s$ parameterizes the relative strength of mutation and selection \citep{haigh:1978}. An example of this distribution is shown in \fig{fig:strong-selection-diagram}. As long as \eq{eq:fitness-distribution} provides a good approximation to the actual stochastic class sizes, the corresponding patterns of diversity are equivalent to a demographically structured \emph{neutral} population, where the $h_k$ are treated as fixed subpopulations and deleterious mutations are recast as migration between the $h_k$'s (see \app{appendix:structured-coalescent}). This is a special case of the \emph{structured coalescent} introduced by \citet{kaplan:etal:1988}, which traces the ancestry of a sample as it moves through the population fitness distribution (see \fig{fig:strong-selection-diagram}).

This simplified structured coalescent admits approximate analytical calculations for several simple diversity statistics.  These reduce to the standard background selection\footnote{There is some ambiguity in the literature regarding the term ``background selection,'' specifically whether it refers to the general effects of purifying selection at linked neutral loci or to the limiting behavior that arises when selection is extremely strong. Here, we use the term in the latter sense as defined by \eq{eq:background-selection}} limit when $\Ns \to \infty$ \citep{charlesworth:etal:1993}, in which the linked neutral diversity is equivalent to an unstructured neutral population with effective population size
\begin{align}
N_e \approx N e^{-\lambda} \, .
\label{eq:background-selection}
\end{align}
Non-neutral corrections arise for $Ns < \infty$ and require more complicated calculations \citep{gordo:etal:2002, walczak:etal:2012,nicolaisen:desai:2012}. More importantly in practice, the simplified structured coalescent can be used to efficiently simulate genealogies in a time that scales with the size of the sample rather than the size of the population \citep{hudson:kaplan:1994,gordo:etal:2002}. These coalescent simulations can rapidly generate predictions for many statistics of interest, and could potentially be leveraged to enable full-scale inference of population parameters.

However, this simplified picture is crucially dependent on the assumption that the fitness class sizes (and hence the mutation and coalescence rates) are effectively fixed by the deterministic mutation-selection balance in \eq{eq:fitness-distribution}. In general, genetic drift will cause the actual class sizes to fluctuate around these deterministic predictions, so the validity of our assumption will depend on the severity of these fluctuations. Given our assumption of one-way mutation, there is also a nonzero probability that the least-loaded ($k=0$) class fluctuates to extinction, allowing one of the deleterious alleles to fix. This effect is known as \emph{Muller's ratchet}, and it implies that the deterministic mutation-selection balance is stochastically unstable. In the absence of compensatory forces, the entire population will tend to drift toward lower fitness at some small but nonzero rate.

While it is not surprising that these stochastic forces eventually cause significant deviations from \eq{eq:fitness-distribution}, a quantitative characterization of this breakdown is complicated, and the precise answer depends on the quantity of interest. For example, one could examine fluctuations in the fitness class sizes \citep{neher:shraiman:2012}, the transition between the so-called ``slow" and ``fast" regimes of Muller's ratchet \citep{gessler:1995}, the breakdown of the background-selection limit \cite{gordo:etal:2002}, or empirical estimates of divergence from the structured coalescent  (see below). Fortunately, most of these definitions of ``strong selection" lead to conditions of the form $\Ns e^{-\lambda} \gg g(\lambda)$, where $g(\lambda)$ is some slowly growing function of $\lambda$. In practice, the simplified condition
\begin{equation}
\Ns e^{-\lambda} \gg 1
\end{equation}
is generally sufficient to ensure the validity of the structured coalescent for most quantities of interest. We note, however, that even for selection pressures that are traditionally considered to be strong ($\Ns \sim 10$), this condition will be violated if the mutation rates are high enough that many of these mutations segregate in the population at the same time $(\lambda \gg 1)$. Thus, the strong-selection regime could be more accurately described as a strong-selection/weak-mutation regime.

\subsection{Equivalence Principle for Weak Selection}

As selection grows weaker, the deterministic mutation-selection balance provides an increasingly poor estimate of the distribution of fitnesses within the population, as stochastic fluctuations and Muller's ratchet take on a larger role. It is instructive to consider the extreme limit where $\Ns = 0$, when these stochastic forces are strongest. Although we do not normally visualize a neutral population in this way, we could also partition it into ``fitness classes" according to the number of mutations in each individual. In this case, the resulting fitness classes fluctuate wildly on coalescent timescales, and Muller's ratchet ``clicks" at rate $\Ud$.

But if $\Ns$ is identically zero, this population should also be described by the standard neutral coalescent, which ignores all of these complicated factors. Instead of explicitly tracking the number of mutations in each individual, the neutral coalescent places the entire population within a single fitness class of size $N$, where fluctuations can be neglected. This simplification arises because in a neutral population, it does not matter which mutations are accounted for by the fitness classes and which accumulate \emph{within} a fitness class, so long as the total mutation rate is preserved. A population with ``weakly selected'' deleterious mutations of effect $s=0$ is equivalent to one with ``strong selection'' ($\sprime > 0$) if we simultaneously take $\Udprime \to 0$ and $\Unprime \to \Un + \Ud$.

\begin{figure}[t]
\centering
\includegraphics[width=0.95\columnwidth]{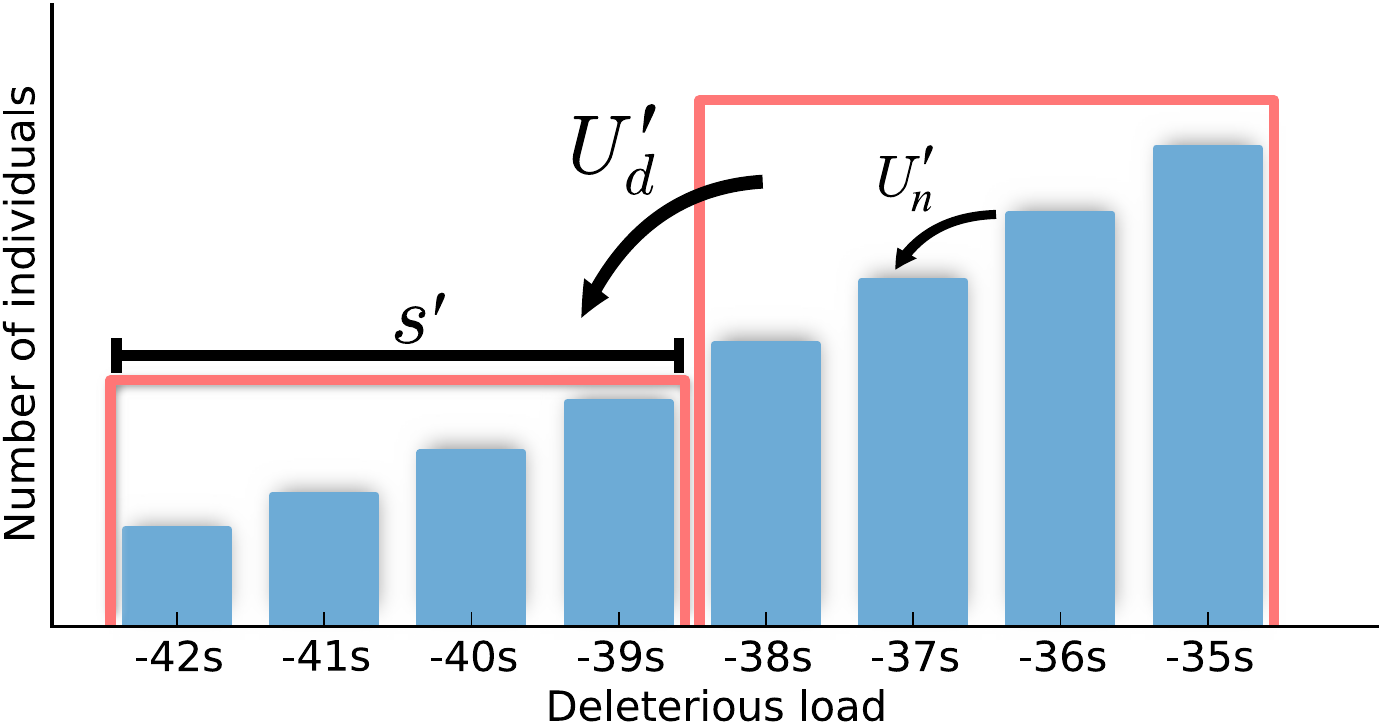}
\vspace{1em}
\caption{An intuitive picture of the coarse-graining proposed for weakly-selected populations, where the population fitness distribution (only a portion of which is shown here) consists of a large number of fitness classes whose sizes fluctuate considerably. Clusters of several fitness classes are grouped together into larger, \emph{effective} fitness classes separated by fitness $\seff$. Deleterious mutations within an effective fitness class are recast as neutral mutations, and only mutations between effective classes are treated as deleterious.  \label{fig:coarse-grained-diagram}}
\end{figure}

This correspondence between weak and strong selection in the neutral limit is admittedly rather trivial, but it suggests that a similar reorganization may hold more generally for weak but non-vanishing $\Ns$. Intuitively, we seek a coarse-grained version of the fitness distribution at some larger scale $\sprime$, such that fitness differences less than $\sprime$ are ignored, but clusters of mutations with cumulative effect $\sprime$ are treated as a single, large-effect mutation (see \fig{fig:coarse-grained-diagram}). In addition, we wish to choose $\sprime$ and the reorganized mutation rates $\Udprime$ and $\Unprime$ in order to mimic (as closely as possible) the patterns of diversity in the original population. It is not clear that this equivalent population should exist \emph{a priori}, and even if it does, the new parameters $\Nsprime$, $\NUdprime$, and $\NUnprime$ could potentially depend on the underlying parameters $\Ns$, $\NUd$, and $\NUn$ in some complicated way. Nevertheless, we demonstrate below that an explicit equivalence principle can be obtained from a few simple considerations.

In our neutral example above, we saw that populations with $\Ns=0$ and various combinations of $\Ud$ and $\Un$ are equivalent as long as the total mutation rate $\NUtot = \NUd + \NUn$ is preserved. For populations with $\Ns > 0$, it is reasonable to expect an additional constraint on the overall scale of selection, which was automatically preserved in the neutral case. This scale is not determined by individual selected mutations, but rather by the emergent distribution of fitnesses within the population. Of course, the fitness distribution is difficult to characterize in weakly-selected populations precisely because of the complicated stochastic effects discussed above. And even if a full solution was available, it is unlikely that that the fitness distributions for two different $(\Ns,\NUd)$ combinations would exactly coincide. Fortunately, previous studies suggest that the dominant feature of the fitness distribution in the $\Ns \to 0$ limit is the variance $\sigma^2$ \citep{ofallon:etal:2010, good:desai:2012b}, which gives a measure of the typical reproductive difference between a random pair of individuals. When the full effects of drift are included, the variance in fitness within the population is given by
\begin{equation}
\label{eq:variance}
\sigma^2 = \Ud s \left[ 1 - \frac{R}{\Ud} \right] \, ,
\end{equation}
where $R$ is the deleterious substitution rate \citep{haigh:1978}. We calculate this rate in \app{appendix:variance}, which completely determines $\sigma^2(\Ns,\NUd)$ as a function of the underlying parameters $\Ns$ and $\NUd$.

Thus, we propose an equivalence principle between weakly selected populations, in which the patterns of diversity are equal when
\begin{equation}
\begin{aligned}
\NUn+\NUd & = \NUdprime+\NUnprime \, , \\
\sigma^2(\Ns,\NUd) & = \sigma^2(\Nsprime,\NUdprime) \, .
\end{aligned}
\label{eq:conservation}
\end{equation}
In the simple case where the deterministic approximation $\sigma^2 \approx \Ud s$ is valid, we have $\Udprime = \Ud \left( s / \sprime \right)$, which has a natural interpretation in terms of the coarse-grained fitness distribution depicted in \fig{fig:coarse-grained-diagram}. It is important to note that the equivalence defined by \eq{eq:conservation} is only \emph{approximately} correct, and it is only valid up to some maximum strength of selection where moments of the fitness distribution other than $\sigma^2$ start to become important.  This breakdown is unsurprising: when $\Ns e^{-\lambda} \gg 1$ we must recover the strong selection limit, where it is known that the least-loaded class plays a much larger role than the bulk of the fitness distribution \citep{charlesworth:etal:1993, nicolaisen:desai:2012, neher:shraiman:2012}.

When our equivalence principle holds, \eq{eq:conservation} defines an \emph{equivalence class} of populations indexed by a particular value of $\NUtot$ and $\Nsigma$, in which different underlying parameters $\Ns$, $\NUd$, and $\NUn$ nevertheless generate the same patterns of diversity. Yet the vast majority of these populations lie well beyond the range of validity of existing methods like the structured coalescent. In order to actually \emph{predict} the patterns of diversity in these populations, we therefore look for a representative population $(\Nseff,\NUdeff,\NUdeff)$ within each equivalence class where selection is simultaneously weak enough for our equivalence principle to hold and yet strong enough for these previously developed techniques to be valid.

\begin{figure}[t]
\centering
\includegraphics[width=0.95\columnwidth]{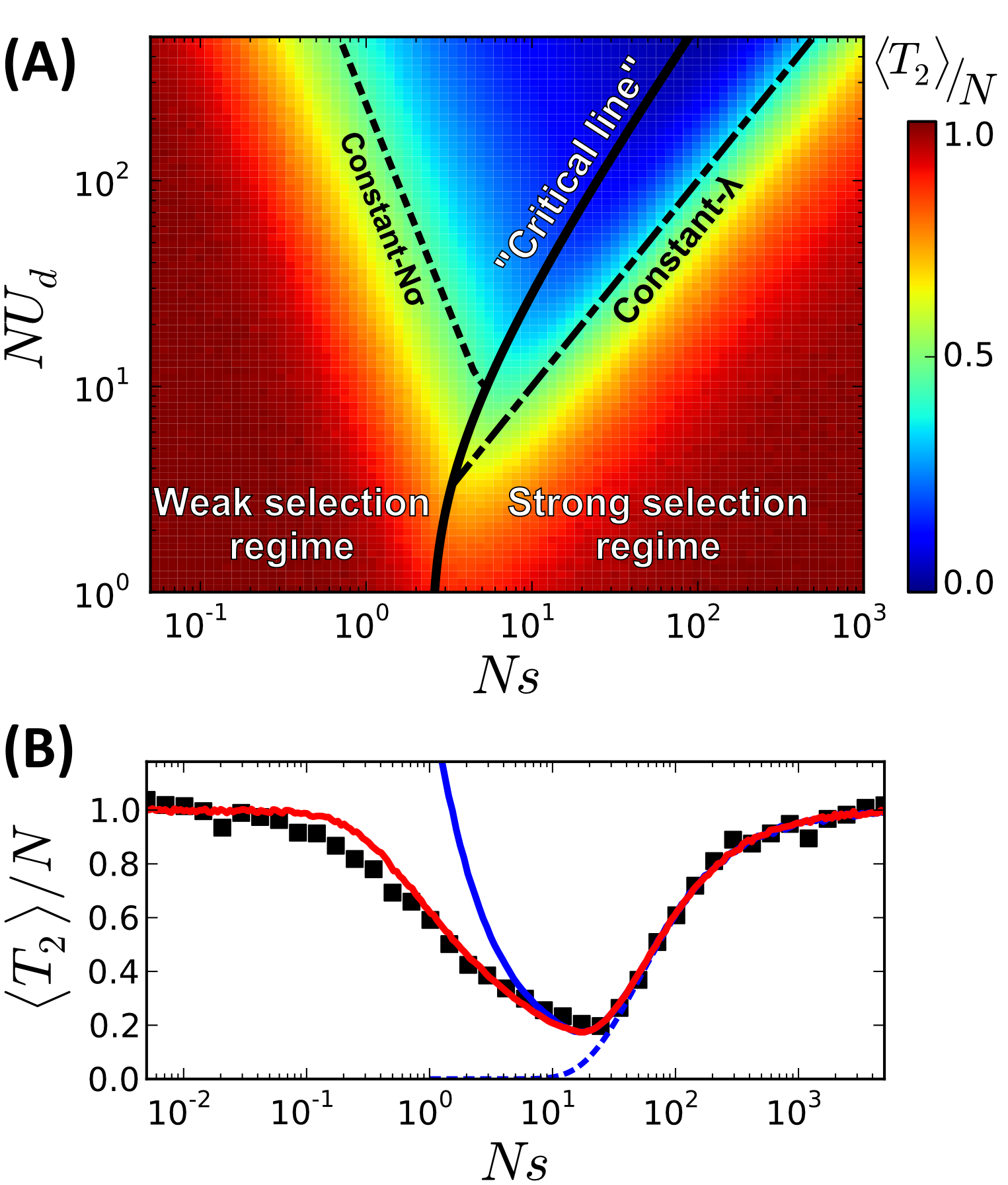}
\vspace{1em}
\caption{(A) Predictions for the mean pairwise coalescence time $\T2/N$ obtained from structured coalescent simulations of our coarse-grained model. The solid black line denotes the ``critical line" $(\Nseff,\NUdeff)$ defined by \eq{eq:critical-line}, while the dashed lines to the left and right denote lines of constant $N\sigma$ and lines of constant $\lambda$, respectively. (B) A ``slice" of this phase plot at constant $\NUd = 50$. The black squares denote the results of forward-time, Wright-Fisher simulations and our coarse-grained predictions are shown in solid red. For comparison, the solid blue line shows the original structured coalescent predictions, while the dashed blue line shows the background selection approximation $\T2 \approx N e^{-\lambda}$. \label{fig:t2}}
\end{figure}

From our numerical and analytical studies of the structured coalescent (see \app{appendix:coarse-grained-model}), we have seen that the predictions for the mean pairwise coalescence time $\langle T_2 \rangle/N$ reach a minimum for a particular value of $\Ns$, below which the predictions rapidly diverge from the results of forward-time simulations (see \fig{fig:t2}b). It therefore seems reasonable to take $\Nseff$ to be this minimum point, which satisfies
\begin{equation}
\label{eq:critical-line}
\left( \frac{\partial \langle T_2 \rangle}{\partial s} \right)_{\sigma^2} = 0 \, .
\end{equation}
The subscript denotes that the fitness variance $\sigma^2$ is to be held constant when taking the derivative. In \app{appendix:coarse-grained-model}, we show how this derivative can be calculated using the methods outlined in \citet{walczak:etal:2012}. The resulting locus of points yields a \emph{critical line} in the $(\Ns,\NUd)$ plane parameterized by the fitness variance $(N\sigma)^2$, as depicted in \fig{fig:t2}a. Each point along this critical line corresponds to a coarse-grained model with parameters $(\Nseff, \NUdeff)$ where the structured coalescent is valid, and which can be used to predict the patterns of diversity for all weakly selected populations within that equivalence class. For any particular set of parameters, the corresponding coarse-grained model can be easily calculated from \eqs {eq:conservation}{eq:critical-line}, with the help of our expression for $\sigma^2$ in \eq{eq:variance}.

\section{Results}

In the previous section, we argued that the patterns of molecular diversity should be equivalent for weakly selected populations with the same total mutation rate and variance in fitness. Within each of these equivalence classes, we have also identified a particular ``coarse-grained'' population $(\Nseff,\NUdeff, \NUneff)$ where previous strong selection methods based on the structured coalescent can be applied. This mapping yields explicit predictions for various diversity statistics across the full range of selection strengths, an example of which is shown in \fig{fig:t2}a for the mean pairwise coalescent time $\T2/N$. Parameters that fall to the right of the critical line (depicted by the solid line in \fig{fig:t2}a) lie in the strong selection regime where $\T2/N$ is directly calculated from the structured coalescent. The vast majority of these points are well-characterized by the background selection limit in \eq{eq:background-selection}, which implies that the level sets of $\T2/N$ lie along lines of constant $\lambda$ (the dashed-dotted line in \fig{fig:t2}a). As observed in previous studies, this strong selection equivalence starts to break down near the critical line where the full structured coalescent is required to obtain accurate predictions \citep{gordo:etal:2002,walczak:etal:2012}. Those parameter values that lie to the left of the critical line are the domain of our coarse-grained theory and the corresponding equivalence along lines of constant $N\sigma$ (depicted by the dashed line in \fig{fig:t2}a).  We obtain predictions for these populations by applying the structured coalescent to the coarse-grained parameters $(\Nseff,\NUdeff,\NUneff)$ calculated using the procedure in \app{appendix:coarse-grained-model}. Intuitively, this amounts to tracing the line of constant $N \sigma$ back to the corresponding point on the critical line in \fig{fig:t2}a.

The accuracy of these predictions rests on two crucial assumptions, which we verify using forward-time, Wright-Fisher simulations (described in \app{appendix:forward-time-simulations}) for several important and experimentally relevant diversity statistics. First, populations with the same fitness variance $(N\sigma)^2$ should yield similar results for various diversity statistics. Secondly, structured coalescent predictions should agree with the results of forward time simulations along the critical line $(\Nseff,\NUdeff)$.

\begin{figure}[t]
\centering
\includegraphics[width=0.95\columnwidth]{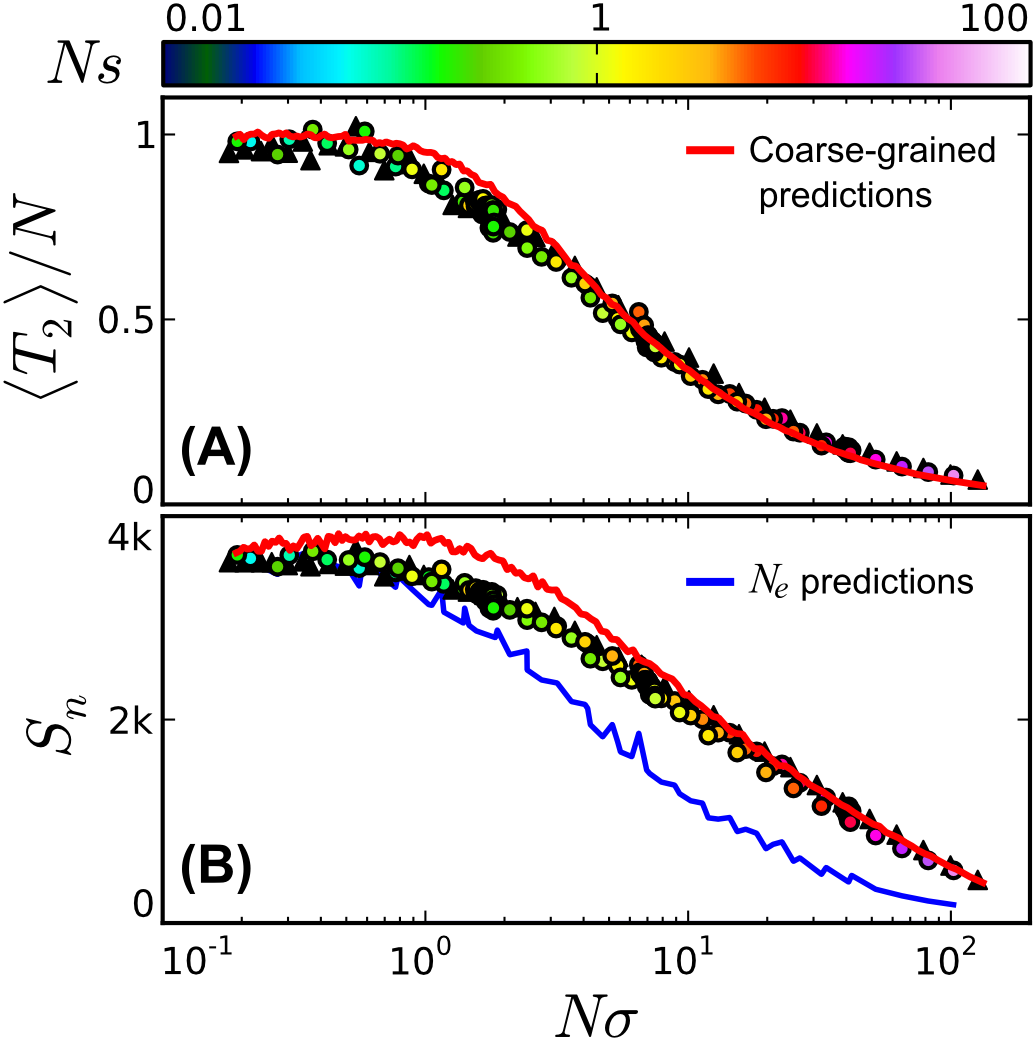}
\vspace{1em}
\caption{(A) The mean pairwise coalescence time $\T2/N$ in the weak-selection regime (top), collated from forward-time, Wright-Fisher simulations at five ``slices'' of constant $\NUd=1,10,50,100$ and $300$ (similar to \fig{fig:t2}b), and the three ``slices'' of constant $N\sigma = 2, 7, 43$ shown in \fig{fig:frequency-collapse}. Each simulated population is plotted according to its fitness variance $\Nsigma$ (averaged over the simulation run) and colored according to the underlying selection strength $\Ns$. In addition, direct forward-time simulations of the critical line $(\Nseff,\NUdeff)$ are shown as black triangles, while the predictions from the structured coalescent are shown in solid red. (B) A similar plot for the total number of segregating sites $\Sn$ in a sample of $n=100$ individuals, where the total mutation rate is given by $\NUtot=350$. For comparison, the blue line shows the predictions obtained by assuming independent evolution at different sites, with an effective population size $\Ne$ fitted from $\T2$ above. \label{fig:collapse}}
\end{figure}

As we demonstrate in \figs{fig:collapse}{fig:frequency-collapse}, both of these assumptions are approximately valid across a large range of parameter values. In \fig{fig:collapse}a, we plot the mean pairwise coalescent time $\T2/N$ obtained from forward-time simulations for a large collection of parameters spanning several orders of magnitude in $\Ns$ and $\NUd$, all of which lie to the left of the critical line in \fig{fig:t2}a. The results are organized by their observed fitness variance along the $x$-axis and colored according to the selection strength of the simulated population. Differently colored points at the same value of $\Nsigma$ represent populations with different underlying parameters that fall within the same predicted equivalence class. If our equivalence principle is correct, these colored points should all lie on the same line. In addition, we also plot the structured coalescent predictions for the coarse-grained parameters $(\Nseff,\NUdeff)$ as a function of $\Nsigma$, which show good agreement with both forward-time simulations of the critical line (black triangles) and the other populations in each equivalence class.

One of the reasons that the pairwise coalescent time $\T2$ plays such a prominent role in earlier studies is that in the standard Hill-Robertson picture, it is equivalent to the effective population size that supposedly captures all of the effects of linked selection. Even when $\T2$ cannot be predicted analytically, it can be measured from the average heterozygosity at putatively neutral (e.g. synonymous) sites. Both the selected and non-selected sites are then assumed to evolve independently with an effective population size $\Ne = \T2$. While this intuition works well for predicting the average nonsynonymous heterozygosity [\citet{kaiser:charlesworth:2008}, see \app{appendix:coarse-grained-model}], several studies have shown that it fails for other statistics that are more sensitive to the correlations produced by linked selection \citep{santiago:cabellero:1998,comeron:kreitman:2002,comeron:etal:2008}. In \fig{fig:collapse}b, we plot the total number of segregating sites $\Sn$ in a sample of $n=100$ individuals in a similar manner as \fig{fig:collapse}a. We see that even after conditioning on the ``correct'' reduction in $\Ne = \T2$, this independent sites assumption significantly underestimates the total diversity in the sample when $\Nsigma > 1$.  By contrast, the structured coalescent predictions of the coarse-grained model yield accurate results for the full range of parameters, without needing to fit to the correct $\T2$.

\begin{figure}[t]
\centering
\includegraphics[width=0.95\columnwidth]{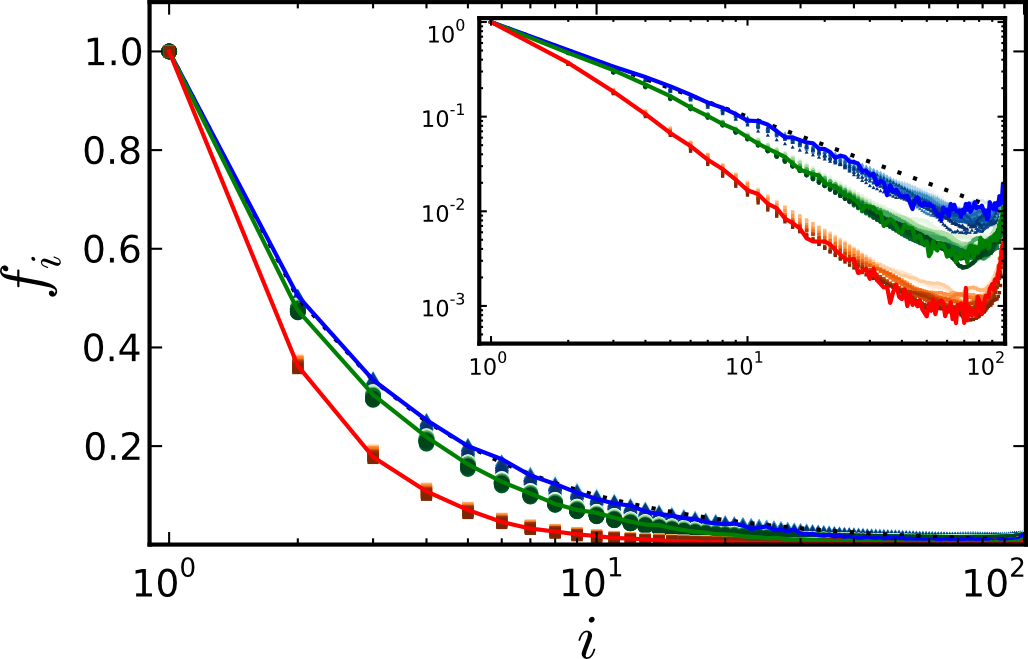}
\vspace{1em}
\caption{The polymorphic site frequency spectrum for a sample of $n=100$ individuals, collated from populations with total mutation rate $\NUtot = 350$ and fitness variance $\Nsigma \approx 2$ (blue), $\Nsigma \approx 7$ (green), and $\Nsigma \approx 43$ (red). Symbols denote the results of forward-time simulations (which are shaded from light to dark with decreasing $\Ns$), while the corresponding structured coalescent predictions are shown as solid lines. For comparison, the prediction for a completely neutral population is depicted by the dashed line. The inset shows the same figure on a log-log scale. \label{fig:frequency-collapse}}
\end{figure}

In addition to reducing the overall levels of diversity described by statistics like $\T2$ and $\Sn$, it is also well-known that purifying selection alters the relative branch lengths in the genealogy of a sample. This distortion is typically measured using the polymorphic site frequency spectrum or one of its derivatives such as Tajima's $D$ \citep{tajima:1989} or Fu and Li's $D$ \citep{fu:li:1993}. When normalized by the total number of singletons, the neutral expectation for the frequencies $f_i$ of the sites polymorphic in $i$ individuals is given by the parameter-free estimate
\begin{equation}
\label{eq:neutral-frequency-spectrum}
f^{\mathrm{neutral}}_i = i^{-1} \, .
\end{equation}
Purifying selection leads to an increase in rare variants ($i \ll n$) compared to this neutral expectation, since deleterious mutations are typically purged before they can drift to appreciable frequencies. Unlike $\T2$ or $\Sn$, the site frequency spectrum can be used to detect deviations from neutrality without requiring previous estimates of population size or mutation rate, and is therefore highly useful in the analysis of real sequence data. In \fig{fig:frequency-collapse}, we plot the site frequency spectrum for a sample of $n=100$ individuals for a range of populations along three particular lines of constant $\Nsigma$ and $\NUtot$, but to the left of the critical line in \fig{fig:t2}a. Again, we see that populations in the same equivalence class possess very similar frequency spectra, which turn increasingly non-neutral with larger $\Nsigma$. We also show the structured coalescent predictions of the coarse-grained model for each value of $\Nsigma$, which agree quite well with these forward-time simulations.

However, despite the generally good agreement with forward-time simulations, some small systematic errors remain. For example, the postulated equivalence between populations with the same fitness variance is only approximately true. As can be seen in \figs{fig:collapse}{fig:frequency-collapse}, populations that are further from the critical line are generally slightly ``less neutral'' than their counterparts closer to the critical line, although these differences are often dwarfed by the variation along the critical line itself. In addition, the accuracy of the structured coalescent along the critical line is diminished within a region with low $\Nseff$ \emph{and} $\NUdeff$, which leads us to slightly overestimate overall levels of diversity in this region. These issues are discussed in more detail in \app{appendix:coarse-grained-model}.

\section{Discussion}

We have demonstrated an approximate symmetry between the patterns of molecular evolution generated by weak and strong purifying selection. Weakly selected mutations have a negligible impact individually, but a sufficiently large number of these mutations will combine to mimic the effects of a single, stronger mutation whose scale is set by the typical fitness differences within the population. This correspondence allows us to import a large body of theory originally developed for the strong selection regime, which we can use to obtain highly accurate predictions across a much broader range of selection strengths than was previously possible.

Our results are consistent with observations from earlier simulation studies of weak selection \citep{mcvean:charlesworth:2000, comeron:kreitman:2002, gordo:etal:2002, kaiser:charlesworth:2008, seger:etal:2010}, but our coarse-grained model offers a radically different perspective on the relevant processes that contribute to molecular evolution in this regime. Previous work has argued that virtually all of the deviations from the traditional background selection limit  can be attributed to the effects of Muller's ratchet \citep{gordo:etal:2002, seger:etal:2010}, as well as to the influence of weak Hill-Robertson interference, where large fluctuations drive weakly selected alleles to intermediate frequencies \citep{mcvean:charlesworth:2000, comeron:kreitman:2002, seger:etal:2010}. In contrast, our coarse-grained theory includes neither of these complications (aside from the corrections to $\sigma^2$) and still captures the quantitative patterns of variation over broad scales. This suggests that variance in ancestral fitness --- not fluctuations or the ratchet --- is the driving force behind the large-scale patterns of diversity. More complicated stochastic effects may be essential for a first-principles account of linked selection, or for more exotic parameter ranges, but they appear to be of secondary importance for the quantities and regimes considered here.

Our finding that strong selection methods can be used across the full range of parameters has important practical implications for the inference of selection pressures and population sizes from DNA sequence data. Several previous studies have uncovered evidence for the role of weak purifying selection in natural populations \citep{barraclough:etal:2007, loewe:charlesworth:2007, betancourt:etal:2009, seger:etal:2010}, but this evidence is limited by the difficulty in obtaining proper estimates of the population size and selection strength without accounting for the influence of selection at neighboring sites. Previous analyses either ignore linkage altogether \citep{hartl:etal:1994, williamson:etal:2005, keightley:eyre-walker:2007, boyko:etal:2008, tamuri:etal:2012} or depend on computationally costly forward-time simulations evaluated over a narrow range of parameters \citep{kaiser:charlesworth:2008, seger:etal:2010, lohmueller:etal:2011}. Estimates obtained from the first method should be treated cautiously, since the results here and elsewhere demonstrate that the independent-sites assumption is drastically violated when selection at linked sites is common. Although these effects can be treated more rigorously in simulations, our present analysis has uncovered an approximate equivalence or \emph{degeneracy} between populations with weak and strong purifying selection. Equivalent populations are distributed along low-dimensional ``ridges'' embedded in the larger space of parameters, which can be easily missed when simulating a discrete set of parameter values. The degeneracy identified here arises in a simple model with only three parameters, and it is likely that the number of degeneracies will only increase in more complicated models which involve many more parameters. Thus, our analysis argues for a degree of caution when interpreting the estimates from simulation studies as well, since it can be difficult to determine whether there are other ``equivalent'' sets of parameters which are equally (or only slightly less) consistent with the data. Ideally, one could combine the knowledge of the degeneracies identified here with the more computationally efficient coalescent simulations to devise a self-consistent inference scheme that utilizes frequency spectrum data, or potentially even phylogenetic reconstruction \cite{rambaut:etal:2008}. A concrete implementation is beyond the scope of the present paper, but this remains an important avenue for future work.

The correspondence between weak and strong selection also suggests an important \emph{qualitative} shift in the interpretation of polymorphism data. It has been known for some time that the effects of linked selection are more complicated than the traditional picture of independent evolution at a reduced effective population size. But without a simple alternative, the concept of a local effective population size --- which influences the efficacy of selection and varies along the genome in accordance with the neutral heterozygosity  --- remains a popular means of interpreting genomic data \citep{charlesworth:2009, grossmann:etal:2011}. In agreement with previous studies, we have provided further evidence that a simple reduction in effective population size does not lead to consistent results for any statistic other than the average heterozygosity $\pi$, even after fitting $\Ne$ to reproduce the observed reduction in neutral diversity. This leads us to question the ultimate utility of the local effective population size, given that it requires a fit in order to describe only one other property of the data.  Rather, our coarse-grained correspondence suggests that a more suitable local quantity is an \emph{effective strength of selection}. It addition to its increased accuracy in capturing the patterns of diversity as demonstrated above, an effective strength of selection is a more natural candidate for a local measure of linked selection, since different parts of the genome are already under varying degrees of selection. 

However, we must be careful when interpreting this local effective selection strength due to the degeneracy in the parameter space discussed above. The accuracy of the collapse plots in \figs{fig:collapse}{fig:frequency-collapse} hints at a fundamental resolution limit for inferring the underlying selection pressures from polymorphism data alone, since there is little statistical power to differentiate a weakly selected population from its coarse-grained counterpart (or indeed, any other weakly selected population with the same overall mutation rate and fitness variance). This degeneracy is especially problematic for detecting selection at individual sites, given that the coarse-grained mapping works by reassigning selection from some mutations to others with minimal impact on the overall patterns of diversity.

Of course, our analysis is based on a highly simplified model, and additional work will be required to extend these results to more biologically realistic scenarios. Depending on the particular parameter regime involved, epistasis \citep{kimura:maruyama:1966}, finite-site effects \citep{desai:plotkin:2008}, and the presence of beneficial or compensatory mutations \citep{goyal:etal:2012} may all play a larger role than we have assumed here. Particularly questionable is our assumption that all deleterious mutations have the same strength, since it is known that deleterious mutations have a wide distribution of fitness effects in many organisms \citep{eyre-walker:keightley:2007}. Nevertheless, the issues raised here are likely to be a factor in \emph{any} model that includes a sufficiently large number of weakly-selected mutations. At present, there are no analytical descriptions of most of these more complicated scenarios even in the strong-selection regime. Thus while a coarse-grained model of weak selection could be defined in many of these cases, we must first understand the corresponding strong-selection model before coarse-graining can provide useful analytical predictions. 

Possibly more problematic for immediate data analysis is our neglect of recombination in the history of the sample, which limits the direct applications of our theory to asexual organisms, or to mitochondrial DNA or non-recombining regions of the genome in sexual populations. The qualitative issues here remain important in the presence of recombination, since the diversity at each site is influenced by the aggregate selection within some non-recombining neighborhood around it.  However, a quantitative extension of these ideas is difficult, since selection and recombination jointly determine the typical linkage scale and the resulting selection regime within that region \citep{comeron:kreitman:2002, kaiser:charlesworth:2008, mcvean:charlesworth:2000}. We note however that recent work by \citet{zeng:charlesworth:2011} has incorporated finite but nonzero recombination rates into the structured coalescent framework.  Thus when an analogous structured coalescent can be defined, our coarse-graining picture will likely be useful for understanding weak selection in these regimes as well.  This remains an important avenue for future work.

\begin{acknowledgments}
We thank Richard Neher, Aleksandra Walczak, John Wakeley, Oskar Hallatschek, Lauren Nicolaisen, and Elizabeth Jerison for useful discussions. This work was supported in part by the James S. McDonnell Foundation, the Alfred P. Sloan Foundation, and the Harvard Milton Fund. B.H.G. acknolwedges support from a National Science Foundation Graduate Research Fellowship. Simulations in this paper were performed on the Odyssey cluster supported by the Research Computing Group at Harvard University.
\end{acknowledgments}

\bibliographystyle{cbe}
\bibliography{coalescent}

\begin{appendix}

\section{The Structured Coalescent}
\label{appendix:structured-coalescent}

Our theoretical predictions for the diversity statistics in the main text are calculated within the structured coalescent framework, which provides an explicit probabilistic model of the genealogy of a sample from the population. In its most general form, the structured coalescent extends the neutral Kingman coalescent to incorporate arbitrary time-dependent (and possibly stochastic) demographic structure, but this has proven difficult to implement in practice. In the present work, we therefore focus our attention on a particularly simple special case, where the relevant demographic structure is the division of the population into constant fitness classes attained at mutation-selection balance \cite{hudson:kaplan:1994}. As such, this simplified structured coalescent is fundamentally a strong selection result.

In the analysis that follows, we work in the standard coalescent limit $N \to \infty$, where the scaled parameters $\Ns$, $\NUd$, and $\NUn$ are sufficient to characterize the population, and we measure time in units of $N$ generations. As mentioned above, we assume that the distribution of fitnesses within the population is given by the deterministic mutation-selection balance in \eq{eq:fitness-distribution} in the main text, and we neglect fluctuations in the class sizes.

We wish to characterize the possible genealogical histories of a sample of $n$ individuals drawn from the population. In a random sample, these individuals come from fitness classes $k_1,\ldots,k_n$ drawn from the population fitness distribution, which implies that
\begin{equation}
k_1,\ldots,k_n \, \overset{i.i.d.}{\sim} \, \mathrm{Poisson} \left(\lambda = \NUd/\Ns \right) \, .
\end{equation}
We then trace the genealogy back to the most recent common ancestor of the sample. At any given instant, three types of ancestral events can occur:
\begin{enumerate}
\item An individual can experience a neutral mutation at rate
\begin{equation}
N \cdot \left( \frac{N\Un}{N} \right) = N \Un \, .
\end{equation}
\item An individual in class $k_i > 0$ can experience a deleterious mutation (thus transferring it to class $k_i-1$) at rate
\begin{equation}
N \cdot \left(\frac{\Ud N h_{k_i-1}}{N h_{k_i}} \right) = N s k_i
\end{equation}
\item Two individuals in the same fitness class $k = k_i = k_j$ can coalesce to a single individual at rate
\begin{equation}
N \cdot \left( N h_k \right) \cdot \left( \frac{1}{Nh_k} \right)^2 = \frac{1}{h_k}
\end{equation}
\end{enumerate}
These events are competing Poisson processes, which implies that the time to the next event is exponentially distributed with mean equal to the reciprocal of the sum of the rates of all possible events. The event itself is then drawn randomly from the pool of possible events, each weighted by its corresponding rate. This process continues until the sample has coalesced into a single lineage, which is the most recent common ancestor of the sample.

Thus, for a given set of parameters $\Ns$, $\NUd$, and $N\Un$, the distribution of genealogies and mutation events for a sample of $n$ individuals is completely specified. The distribution of any particular diversity statistic can be straightforwardly obtained by averaging over the distribution of genealogies. In practice, evaluating these averages analytically can be difficult for all but the simplest statistics \cite{walczak:etal:2012}. Instead, it is often easier to use the procedure outlined above to implement backward-in-time simulations that sample genealogies from this ancestral process \cite{gordo:etal:2002}. These coalescent simulations are extremely computationally efficient compared to their ordinary forward-time counterparts, since we only need to simulate ancestral events for the sample as opposed to simulating the entire population at each generation. A copy of our implementation in C is available upon request.

\section{Fitness variance under weak selection}
\label{appendix:variance}

In the main text, we postulate an equivalence principle between weakly selected populations with the same variance in fitness, which we calculate here. When selection is strong, $\sigma^2$ can easily be calculated from the mutation-selection balance in \eq{eq:fitness-distribution} in the main text, and we find that
\begin{equation}
\label{eq:strong-selection-variance}
\sigma^2 = \Ud s \, .
\end{equation}
However, we wish to apply these results precisely in the region where the deterministic mutation-selection balance becomes unreliable, so we would like to generalize this calculation to include the full effects of drift as $\Ns \to 0$. A standard calculation shows that more generally, we have
\begin{align}
\sigma^2 = \Ud s \left( 1 - \frac{R}{\Ud} \right) \, ,
\label{eq:variance-identity}
\end{align}
where $R$ is the substitution rate of deleterious mutations \citep{haigh:1978, higgs:woodcock:1995, etheridge:etal:2007, good:desai:2012b}. This reduces to the ordinary strong-selection result in \eq{eq:strong-selection-variance} when $\Ns e^{-\lambda} \gg 1$, but we now focus on the regime where $R$ is not necessarily small compared to $\Ud$. Fortunately, when $R$ is on the order of $\Ud$, this substitution rate is equivalent to the rate of Muller's ratchet in the so-called ``fast-ratchet" regime \citep{gessler:1995}, where many of the complicated stochastic aspects of the ratchet analyzed in \citet{neher:shraiman:2012} can be neglected and results from traveling wave theory \citep{rouzine:etal:2008, hallatschek:2011, good:etal:2012} can be applied.

We calculate the rate of deleterious substitutions using the tunable constraint framework introduced in \citet{hallatschek:2011}, which modifies the standard Wright-Fisher stochastic dynamics in order to make $R$ easier to calculate. These predictions for the substition rate have been shown to agree with ordinary Wright-Fisher simulations in several regimes of positive selection \citep{hallatschek:2011, good:etal:2012}, but they have yet to be directly applied to the purifying selection regime studied here.

We introduce two new quantities $f(x)$ and $w(x)$, which respectively correspond to the population density and the fixation probability of new mutants at relative fitness $x$. In the tunable constraint framework, these are related to the population size and substitution rate through the system of equations
\begin{equation}
\begin{aligned}
s R \partial_x f(x) & = x f(x) - f(x) w(x) \\
	& \quad + \Ud \left[ f(x+s) - f(x) \right] \, , \end{aligned}
\label{eq:tunable-f}
\end{equation}
\begin{equation}
\begin{aligned}
- s R \partial_x w(x) & = x w(x) - w(x)^2 \\
	& \quad + \Ud \left[ w(x-s) - w(x) \right] \, ,
\end{aligned}
\label{eq:tunable-w}
\end{equation}
and the normalization conditions
\begin{equation}
1  = \int_{-\infty}^{\infty} f(x) \, dx \, , \quad \quad \frac{\nu}{N} = \int_{-\infty}^{\infty} f(x) w(x) \, dx \, ,  \label{eq:tunable-constraint}
\end{equation}
where $\nu$ is the variance in offspring number (equal to unity in the Wright-Fisher model). This system of equations can in principle be solved to obtain $R$ as a function of $N$, $s$, and $\Ud$, but in their current form, these equations are difficult to solve (even numerically) due the delay terms $f(x+s)$ and $w(x-s)$ in the differential equations. Thus, we turn to an approximate solution.

Since we are focused on a regime where selection is weak, it seems reasonable to try a Taylor expansion of these delay terms in powers of $s$:
\begin{equation}
\label{eq:mutational-diffusion}
f(x+s) = f(x) + s \partial_x f(x) + \frac{1}{2} \partial^2_x f(x) + \ldots
\end{equation}
In order to obtain a non-trivial solution for $w(x)$, we must include at least the second order term in this expansion, after which \eqs{eq:tunable-f}{eq:tunable-w} can be rewritten in the form
\begin{equation}
\begin{aligned}
0 & = x f(x) - w(x) f(x) \\
	& \quad + \left( \frac{\Ud s^2}{2} \right) \partial^2_x f(x) + \Ud s \left[ 1 - \frac{R}{\Ud} \right] \partial_x f(x) \, ,
\end{aligned}
\end{equation}
\begin{equation}
\begin{aligned}
0 & = x w(x) - w(x)^2 \\
	& \quad + \left( \frac{\Ud s^2}{2} \right) \partial^2_x w(x) - \Ud s \left[ 1 - \frac{R}{\Ud} \right] \partial_x w(x) \, .
\end{aligned}
\end{equation}
Note that this is essentially the same model analyzed in \cite{hallatschek:2011}, with the parameters $D$ and $v$ of that work corresponding to $\Ud s^2 / 2$ and $\Ud s ( 1 - R/\Ud)$ here. A similar analysis can therefore be applied in the present case. We first rescale the relative fitness $x$ by introducing the new coordinate
\begin{equation}
\chi = x \left( \frac{\Ud s^2}{2} \right)^{-1/3} \, ,
\end{equation}
Since $w(x)$ and $f(x)$ currently have the units of fitness and inverse fitness, respectively, we must rescale these as well:
\begin{equation}
\begin{aligned}
\ftilde(x) & = \left( \frac{\Ud s^2}{2} \right)^{1/3} f(x) \, , \\
\wtilde(x) & = \left( \frac{\Ud s^2}{2} \right)^{-1/3} w(x) \, .
\end{aligned}
\end{equation}
In terms of these rescaled variables, our system of equations can be written in the compact form
\begin{align}
0 & = \partial_{\chi}^2 \ftilde +  \alpha \partial_{\chi} \ftilde + \chi \ftilde - \ftilde \wtilde \, , \label{eq:rescaled-f} \\
0 & = \partial_{\chi}^2 \wtilde - \alpha \partial_{\chi} \wtilde + \chi \wtilde - \wtilde^2 \, , \label{eq:rescaled-w} \\
1 & = \int_{-\infty}^{\infty} \ftilde(\chi) \, d\chi \, , \\
\frac{1}{\beta} & = \int_{-\infty}^{\infty} \ftilde(\chi) \wtilde(\chi) \, d\chi \, ,
\end{align}
where we have introduced the two parameters
\begin{equation}
\alpha = (4 \lambda)^{1/3} \left( 1 - \frac{R}{\Ud} \right) \, , \quad \beta = \frac{\Ns}{\nu} \left( \frac{\lambda}{2} \right)^{1/3} \, .
\end{equation}
From inspection, we can immediately see that
\begin{equation}
\ftilde(\chi) \propto e^{-\alpha \chi} \wtilde(\chi)
\end{equation}
is a solution to \eq{eq:rescaled-f}, which allows us to eliminate $\ftilde$ entirely and yields the simplified system
\begin{equation}
\begin{aligned}
0 & = \partial_{\chi}^2 \wtilde - \alpha \partial_{\chi} \wtilde + \chi \wtilde - \wtilde^2 \, , \\
\beta & = \frac{\int_{-\infty}^\infty e^{-\alpha \chi} \wtilde(\chi) \, d\chi}{\int_{-\infty}^\infty e^{-\alpha \chi} \wtilde(\chi)^2 \, d\chi} \, ,
\end{aligned}
 \label{eq:simplified-system}
\end{equation}
where $\wtilde(\chi)$ is subject to the boundary conditions $\wtilde(\chi) \to 0$ as $\chi \to -\infty$ and $\wtilde(\chi) \to \chi$ as $\chi \to \infty$. Thus, numerical solution of the boundary value problem (for instance, using Matlab's bvp4c function) and subsequent numerical integration of the resulting $\wtilde$ allows us to calculate $\beta$ as a function of $\alpha$,
\begin{equation}
\beta = g(\alpha) \, ,
\end{equation}
where $g$ is independent of any of the evolutionary parameters. A subsequent inversion of this relation yields an expression for $1-R/\Ud$ as a function of $Ns$ and $\lambda$:
\begin{equation}
\label{eq:ratchet-rate-uncorrected}
1 - \frac{R}{\Ud} = \left( \frac{1}{4 \lambda} \right)^{1/3} g^{-1} \left[ \frac{\Ns}{\nu} \left( \frac{\lambda}{2} \right)^{1/3} \right] \, .
\end{equation}

\begin{figure}[t]
\centering
\includegraphics[width=0.95\columnwidth]{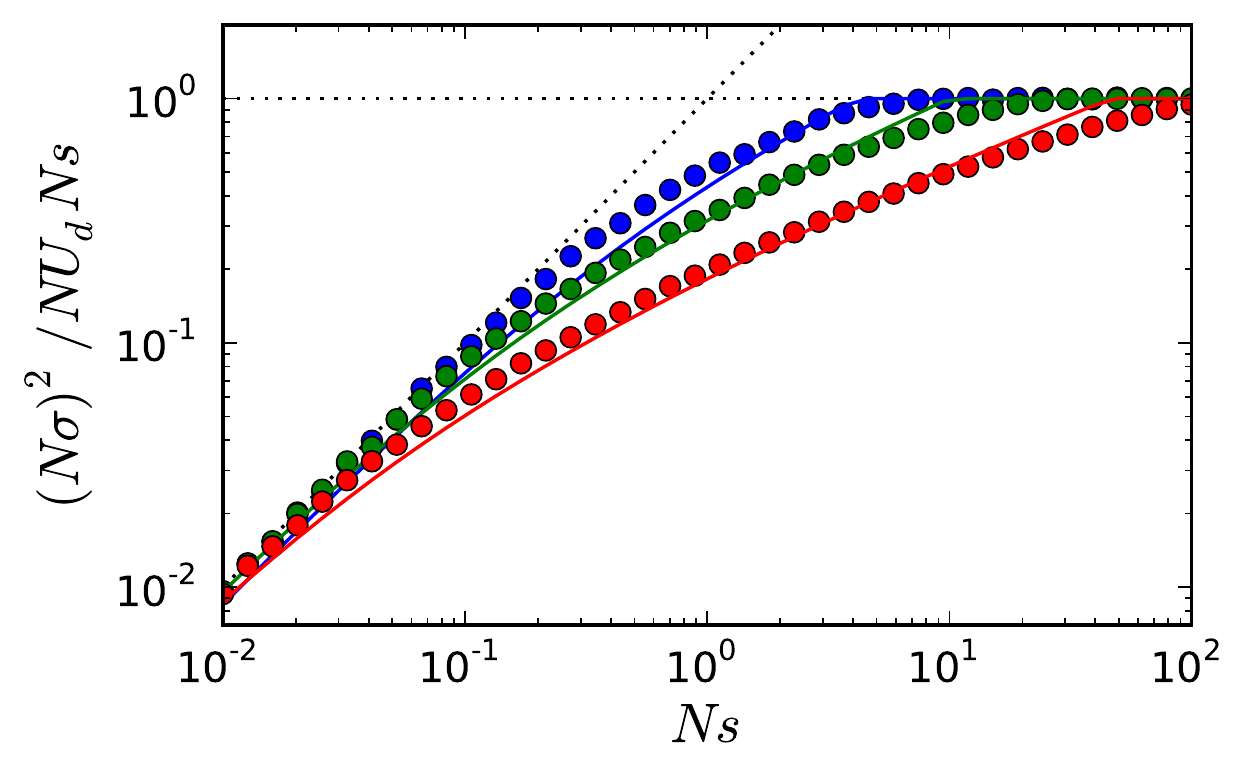}
\caption{The variance in fitness within the population as a function of the selection strength, for $\NUd=5$ (blue), $\NUd=50$ (green), and $\NUd = 500$ (red). Symbols denote the results of forward-time simulations with $N=10^5$ averaged over $300$ independent runs, while the solid lines are the predictions obtained from \eqs{eq:variance-identity}{eq:ratchet-rate}. The dashed lines give the asymptotic behavior for $\Ns \to 0$ and $\Ns \to \infty$. \label{fig:variance}}
\end{figure}

For general values of $\alpha$, this function $g(\alpha)$ must be calculated numerically using the approach outlined above (an implementation in Matlab is available upon request). However, we can obtain simple analytical formulae in two limiting cases. The limit of large $\alpha$ has been studied extensively in previous work \cite{tsimring:etal:1996,hallatschek:2011,goyal:etal:2012}, and we find that $\log g(\alpha) \sim \alpha^3$. In the limit $\alpha \to 0$, the differential equation for $\wtilde$ reduces to the parameter-free form
\begin{equation}
0 = \partial^2_\chi \wtilde + \chi \wtilde - \wtilde^2 \, ,
\end{equation}
whose solution can be reasonably well-approximated in this limit by
\begin{equation}
\wtilde(\chi) \approx \begin{cases}
\chi & \text{if $\chi > 0$,} \\
0 & \text{else.}
\end{cases}
\end{equation}
The normalization integrals in \eq{eq:simplified-system} can then therefore be approximated by $\Gamma$-functions, and we find that $g(\alpha) \approx \frac{\alpha}{2}$. In terms of the underlying evolutionary parameters, this implies that
\begin{equation}
\sigma^2 \approx \left( \frac{\NUd}{\nu} \right) s^2
\end{equation}
as $\Ns \to 0$, which agrees with an analogous calculation using the neutral coalescent.

As a slight technical aside, we note that the left-hand side of \eq{eq:ratchet-rate-uncorrected} by definition cannot be greater than one, since the deleterious mutations cannot accumulate at a negative rate. Nevertheless, as $\Ns$ increases the right-hand side of \eq{eq:ratchet-rate-uncorrected} eventually becomes larger than one (particularly in the large $\alpha$ limit discussed above). This likely indicates a breakdown of the mutational diffusion approximation we assumed in \eq{eq:mutational-diffusion}, or possibly a breakdown in the applicability of the tunable constraint framework in general. In order to maintain sensible results, we therefore take
\begin{equation}
\label{eq:ratchet-rate}
1 - \frac{R}{\Ud} = \mathrm{min} \left\{ \left( \frac{1}{4 \lambda} \right)^{1/3} g^{-1} \left[ \Ns \left( \frac{\lambda}{2} \right)^{1/3} \right] , \, 1 \right\} \, .
\end{equation}

In \fig{fig:variance}, we compare these predictions with the results of forward-time simulations for a representative sample of parameters. The agreement is generally quite good (certainly better than the naive asymptotics alone), although there are some small systematic disagreements. In particular, we tend to slightly overestimate the variance in fitness at the point where it starts to deviate from the deterministic asymptote $\sigma^2 = \Ud s$, and we tend to slightly underestimate it during the transition to the neutral asymptote $\sigma^2 = N \Ud s^2$.

\section{The Coarse-Grained Model}
\label{appendix:coarse-grained-model}

\noindent In the main text, we proposed a theory of weak purifying selection, which was based on two underlying assumptions:
\begin{enumerate}
\item Weakly selected populations with the same $\NUtot$ and $\Nsigma$ form an equivalence class in terms of the patterns of diversity they contain.
\item Within this equivalence class, there exists a population with parameters $(\Nseff,\NUdeff,\NUneff)$ where the strength of selection is strong enough that the structured coalescent is valid.
\end{enumerate}
This allows us to generate predictions for any weakly selected population by identifying the corresponding ``coarse-grained'' population and applying the structured coalescent.

\subsection{Finding the coarse-grained parameters}

For a particular population with parameters $\Ns$, $\NUd$, and $\NUn$, we first identify the corresponding equivalence class by calculating $\Nsigma$ using the formulae in \app{appendix:variance}. Then, as described in the main text, the corresponding coarse-grained population can be found by minimizing $\T2/N$ as a function of $\Ns$, with $\Nsigma$ held constant. This coarse-grained population is by definition in the strong selection regime, so $\T2/N$ can be calculated analytically using the results in \cite{walczak:etal:2012}, which state that
\begin{align}
\label{eq:t2}
\frac{\T2}{N} & = \sum_{k_1=0}^\infty \sum_{k_2 = k_1}^\infty \sum_{k_c = 0}^{k_1} t(k_1,k_2,k_c) p(k_1,k_2,k_c) \, .
\end{align}
The function $t(k_1,k_2,k_c)$ is the mean coalescent time for two individuals in classes $k_1$ and $k_2$ that coalesce in class $k_c$, and is given by
\begin{align}
t(k_1,k_2,k_c) = \frac{h_{k_c}}{1+2 N s k_c h_{k_c}}
\end{align}
in the special case that $k_1 = k_2 = k_c$ and
\begin{equation}
\begin{aligned}
t(k_1,k_2,k_c) & = \sum_{j=1}^{k_1+k_2 - 2k_c} \frac{j (-1)^{j-1} }{(j+2 k_c)^2}  {k_1 + k_2 - 2 k_c \choose j} \\
 & \quad \times {k_1 + k_2 \choose 2 k_c} \left[ \frac{1}{\Ns} + \frac{(j+ 2 k_c) h_{k_c}}{1+2 \Ns k_c h_{k_c}} \right]
\end{aligned}
\end{equation}
otherwise. The function $p(k_1,k_2,k_c)$ is the probability of sampling two individuals from classes $k_1$ and $k_2$ that coalesce in class $k_c$, and is given by
\begin{equation}
\begin{aligned}
p(k_1,k_2,k_c) & = \frac{{k_1 \choose k_c} {k_2 \choose k_c} {k_1+k_2 \choose 2 k_c}^{-1} h(k_1,k_2)}{1 + 2 Ns k_c h_{k_c}}   \\
	& \quad \times \prod_{j=k_1}^{k_c-1} \left[ 1 - \frac{{k_1 \choose j} {k_2 \choose j} {k_1+k_2 \choose 2 j}^{-1}}{1 + 2 Ns j h_{j}} \right]
\end{aligned}
\end{equation}
where
\begin{align}
h(k_1,k_2)	= \begin{cases}
h_{k_1}^2 & \text{if $k_1 = k_2$,} \\
2 h_{k_1} h_{k_2} & \text{else.}
\end{cases}
\end{align}
In addition, when \eq{eq:t2} is valid, the fitness variance is related to $\Ns$ and $\NUd$ through the simple relation
\begin{equation}
(\Nsigma)^2 \approx (\NUd) (\Ns) \, .
\end{equation}
Putting these two facts together, we see that $\Nseff$ is given by the root of the equation
\begin{equation}
0 = \frac{\partial \T2}{\partial (Ns)} - \frac{2 \lambda}{\Ns} \frac{\partial \T2}{\partial \lambda} \, ,
\end{equation}
where we set $\lambda = (\Nsigma/\Ns)^2$ after taking the derivative. Once we have determined $\Nseff$, the corresponding mutation rates $\NUdeff$ and $\NUneff$ are given by
\begin{equation}
\begin{aligned}
\NUdeff & = \frac{(\Nsigma)^2}{\Nseff} \, , \\
\NUneff & = \NUn + \left( \NUd - \NUdeff \right) \, .
\end{aligned}
\end{equation}
Thus, we have constructed an explicit mapping between underlying parameters $(\Ns,\NUd,\NUn)$ and those of the corresponding coarse-grained model $(\Nseff,\NUdeff,\NUneff)$ where we can apply the structured coalescent. A copy of our implementation in Python is available upon request.

\subsection{Deviations from the coarse-grained predictions}

This procedure enables us to generate predictions for any set of parameters $\Ns$, $\NUd$, and $\NUn$, but the validity of these predictions depends on the validity of the underlying assumptions (1) and (2) stated above. Figure 3 (main text) shows that these assumptions are \emph{approximately} true over a broad parameter regime, but some small systematic deviations are observed. In order to examine these deviations in more detail, we focus on a narrower (yet still representative) set of parameters.

\begin{figure}[t]
\centering
\includegraphics[width=0.95\columnwidth]{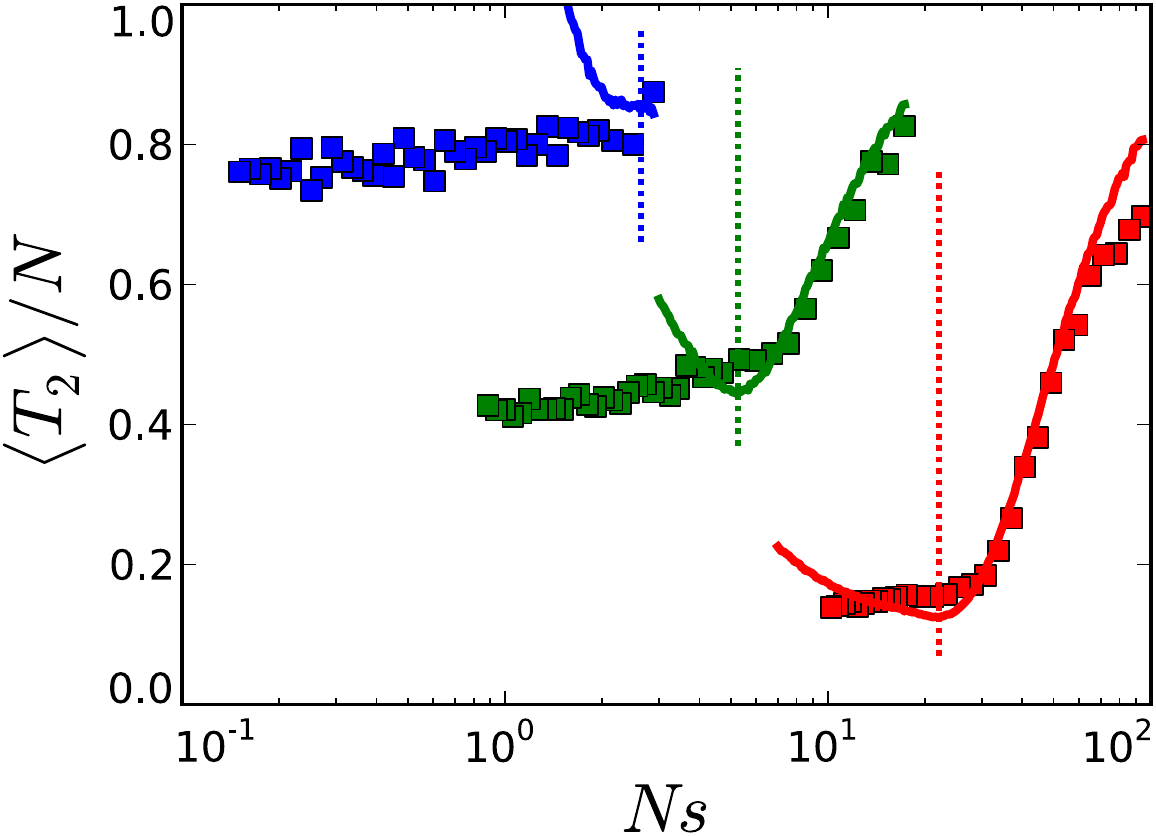}
\caption{The mean pairwise coalescence time as a function of the selection strength, for fixed $\Nsigma \approx 2$ (blue), $\Nsigma \approx 7$ (green), and $\Nsigma \approx 43$ (red). Symbols denote the results of forward-time simulations, while the solid lines show the predictions from the original structured coalescent. The dotted lines denote the corresponding points $(\Nseff,\NUdeff)$ on the critical line, which are utilized by our coarse-grained theory. \label{fig:equivalence-t2}}
\end{figure}

In the same manner as Figure 4 in the main text, we simulate three lines of constant $\Nsigma$ and $\NUtot$ [calculated from \eqs{eq:variance-identity}{eq:ratchet-rate}] for $\Nsigma \approx 2$, $\Nsigma \approx 7$, and $\Nsigma \approx 43$. Results for the mean pairwise coalescent time are shown in \fig{fig:equivalence-t2}. In all three cases, we observe a characteristic ``hockey-stick'' shape, in which a region of rapidly varying $\T2/N$ sharply transitions to a region of significantly reduced variation. This abrupt transition coincides with the minimum of the structured coalescent predictions for $\T2/N$, and hence with the boundary of the strong selection regime.

\begin{figure}[t]
\centering
\includegraphics[width=0.95\columnwidth]{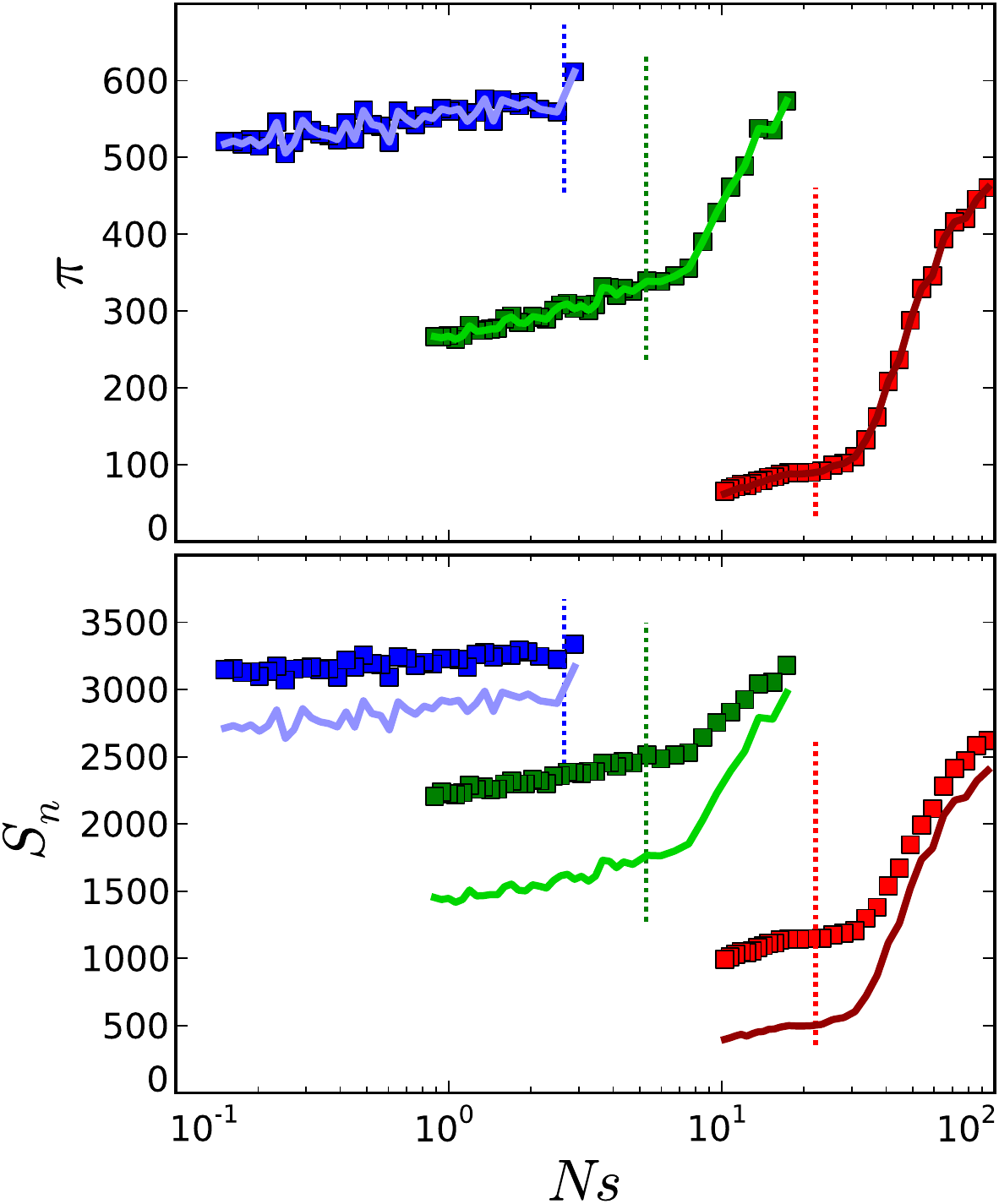}
\caption{The mean pairwise heterozygosity (top) and the total number of segregating sites for a sample of $n=100$ individuals (bottom) as a function of selection strength for the lines of constant $\Nsigma$ shown in \fig{fig:equivalence-t2}, with $\NUtot = 350$. Symbols denote the results of forward-time simulations, while the solid lines show the predictions from the effective population size assumption, fitted to the simulated values of $\T2$ in \fig{fig:equivalence-t2}. The dotted lines denote the corresponding points $(\Nseff,\NUdeff)$ on the critical line, which are utilized by our coarse-grained theory. \label{fig:equivalence-stats}}
\end{figure}

If assumption (1) is exactly satisfied, the points to the left of this boundary should all have the same value. We see that this is true to a good degree of approximation, in the sense that the remaining variation in these points is much less than the variation \emph{between} the lines with different $\Nsigma$. Nevertheless, there remains a slight downward trend along these lines of constant $\Nsigma$ as selection grows weaker, which corresponds to these points being slightly ``less neutral'' than we would predict using our equivalence class. In \fig{fig:equivalence-stats} we plot the mean pairwise heterozygosity and the total number of segregating sites for these lines as well. We observe a similar  ``hockey stick" shape, although the deviation from assumption (1) is slightly stronger than we observed for $\T2/N$.

\begin{figure}[t]
\centering
\includegraphics[width=0.95\columnwidth]{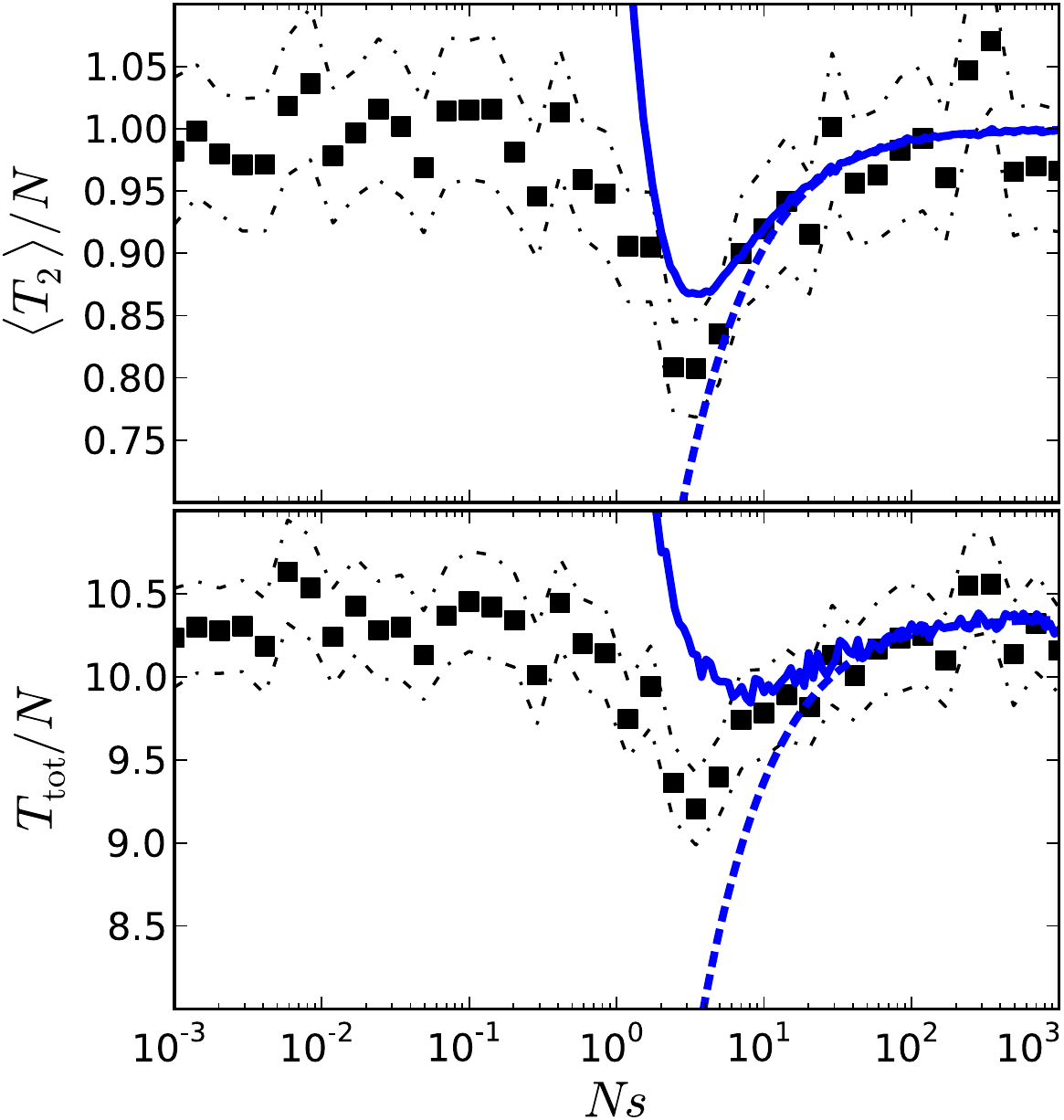}
\caption{The mean pairwise coalescent time (top) and the total tree length for a sample of $n=100$ individuals (bottom) as a function of selection strength, for fixed $\NUd=1$. Symbols denote the results of forward-time simulations, and the black dashed lines give approximate 95\% confidence intervals. The solid blue line shows the predictions from the original structured coalescent, while the dashed blue line is the corresponding background selection approximation. \label{fig:low-mutation}}
\end{figure}

Our coarse-grained theory also depends on the validity of assumption (2), which requires that the structured coalescent predictions along the critical line should match forward-time simulations without any further modifications. Again, while \figs{fig:collapse}{fig:frequency-collapse} in the main text show that this is generally true, we do observe some systematic deviations, especially for points where both $\Nseff$ \emph{and} $\NUdeff$ are small. We can examine this regime more closely in \fig{fig:low-mutation}, which plots the pairwise coalescent time as a function of the selection strength for constant $\NUd=1$.

We observe that at these low values of $\lambda = \NUd/\Ns$, the structured coalescent overestimates the characteristic minimum value of $\T2$, although it gets the location more or less correct. Typically, these low values of $\lambda$ are associated with extremely strong selection pressures, so that the classic background selection approximation is valid. However, we see that near this minimum -- which represents the maximum deviation from neutrality for this level of mutation -- neither background selection nor the structured coalescent gives the correct result. Our coarse-grained theory can therefore do no better.

Further study of this low $\NUd$ and low $\Ns$ region may shed light on the interactions between stochastic fluctuations and the structured coalescent framework, and could offer insight on how to incorporate first order corrections for these effects.  However, these deviations are generally small, and for such low values of $\Ns$ and $\NUd$ the population is nearly neutral anyway. In addition, small discrepancies in this relatively narrow region of parameter space are unlikely to matter much for practical purposes, since the patterns of diversity observed in actual populations are likely to be dominated by larger deviations from neutrality attained at larger values of $\Ns$ and $\NUd$.

\subsection{Comparison with the effective population size picture}

The diversity statistics in \fig{fig:equivalence-stats} allow us to make a more detailed comparison with the effective population size picture that is typically used to interpret such data. In this case, $\Ne$ can be measured exactly by fitting to the corresponding $\T2$ values in \fig{fig:equivalence-t2}, and the predictions for $\pi$ and $\Sn$ (shown as solid lines in \fig{fig:equivalence-stats}) follow from assuming that the individual sites otherwise evolve independently at this reduced effective population size. We see that while this assumption leads to excellent agreement for $\pi$ (as observed previously in \cite{kaiser:charlesworth:2008}), it drastically underestimates $\Sn$, which depends more sensitively on the genealogical distortions caused by purifying selection.

\begin{figure}[t]
\centering
\includegraphics[width=0.95\columnwidth]{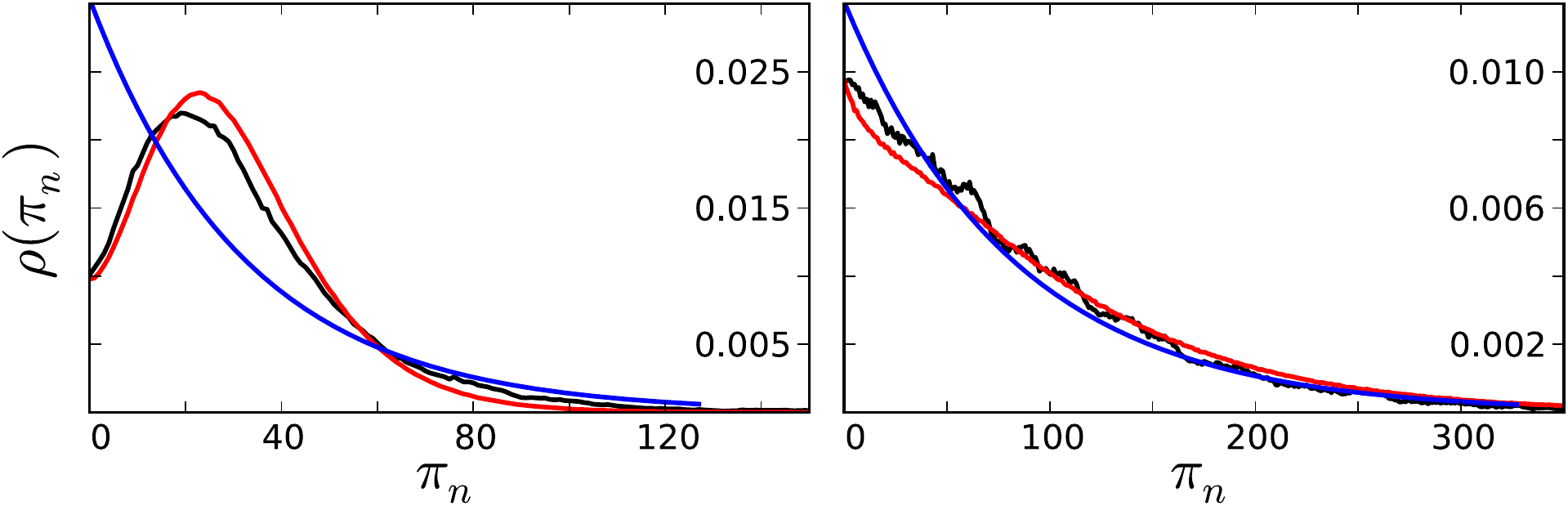}
\caption{The distribution of neutral pairwise heterozygosity for a population with $\Ns = 5$ (left) and $\Ns = 0.25$ (right), with $\NUd = \NUn = 50$. Black lines denote the results of forward-time simulations and the red lines show the structured coalescent predictions for our coarse-grained theory. For comparison, the blue lines show the predictions from the standard effective population size picture, with $\Ne$ fitted from the mean of the forward-time distribution. \label{fig:pin-dist}}
\end{figure}

\begin{figure}[t]
\centering
\includegraphics[width=0.95\columnwidth]{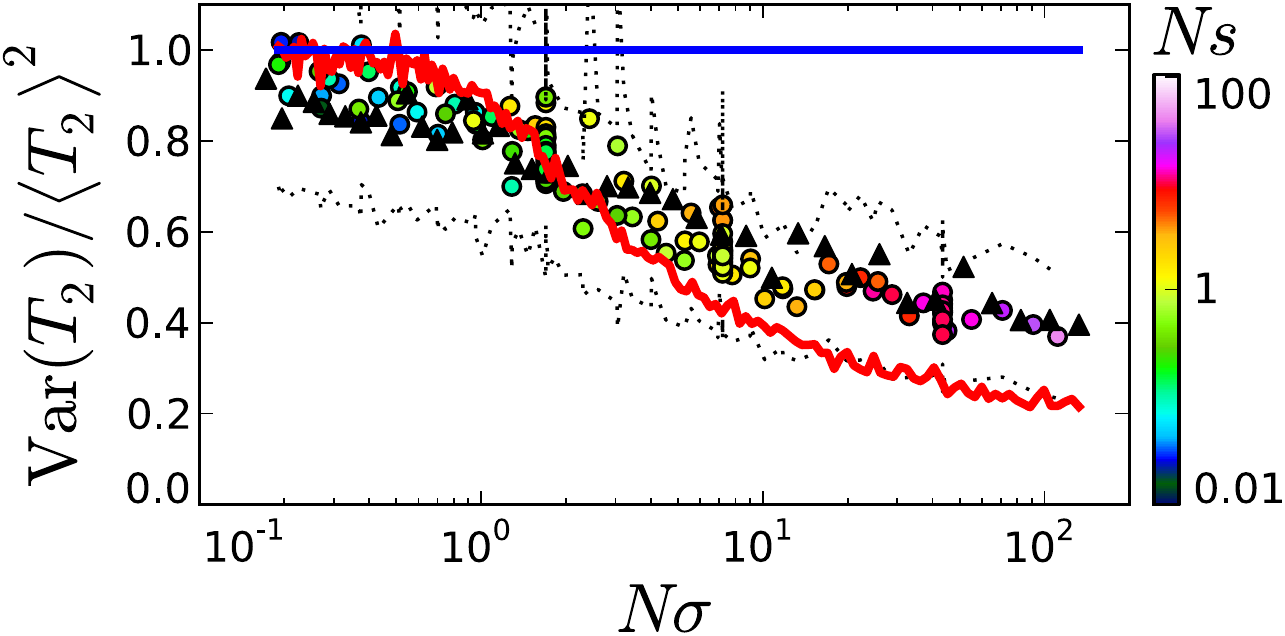}
\caption{The relative variance in pairwise coalescent time as a function of $\Nsigma$ for the same populations as Figure 3 in the main text. Again, symbols denote the results of forward-time simulations (black dotted lines show approximate 95\% confidence intervals), while the solid red line gives the structured coalescent predictions of our coarse-grained theory. For comparison, the predictions from the standard effective population size picture are shown as a solid blue line. \label{fig:t2-variance-collapse}}
\end{figure}

We can find even larger discrepancies by focusing on the \emph{distributions} of certain statistics, --- particularly those related to the neutral diversity --- which take on an extremely simple form in the effective population size picture. For example, under this assumption the distribution of neutral pairwise heterozygosity is predicted to follow a geometric distribution with mean $2\Ne \Un$. The most likely value is $\pi_n = 0$, and the probability of larger values decreases monotonically with increasing $\pi_n$. In \fig{fig:pin-dist}, we plot the distribution of neutral heterozygosity for two populations in the weak selection regime. We see that this effective population size assumption fails to capture the qualitative features of the distribution, despite being fitted to the correct mean value. This effect is exaggerated closer to the critical line, where the distribution develops a strong peak at a nonzero value of $\pi_n$ resulting from a corresponding peak in the distribution of $T_2$.

We can quantify this peaked nature of the distribution over a broader range of parameters by looking at the variance in the pairwise coalescent time. Under the effective population size assumption, $T_2$ follows an exponential distribution with mean $\Ne$. This implies that the ratio of the variance and the mean is given by
\begin{equation}
\frac{\mathrm{Var}(T_2)}{\T2^2} = 1 \, ,
\end{equation}
independent of $\Ne$ or any of the other parameters. In \fig{fig:t2-variance-collapse}, we measure this statistic for each of the populations in \fig{fig:collapse} (main text) and construct an analogous collapse plot. Again, our equivalence principle is highly accurate, and our coarse-grained predictions from the structured coalescent quantitatively describe these effects of linked selection that are not even qualitatively captured by the effective population size picture.

\section{Forward-time Simulations}
\label{appendix:forward-time-simulations}

We validate several of our key approximations in the main text by comparing our theoretical predictions with the results of forward-time, discrete-generation simulations similar to the standard Wright-Fisher model \cite{ewens:2004}. These simulations begin with a clonal population of $N$ individuals, and in each subsequent generation the population undergoes a selection step followed by a mutation step. In the selection step, each lineage (i.e. unique genotype) is assigned a new size from a Poisson distribution with mean
\begin{equation}
\lambda_i = C (1+x_i) n_i  \, ,
\end{equation}
where $n_i$ is the current size of the lineage, $x_i$ is its fitness relative to the population average, and $C = N / \sum_i n_i$ is a normalization constant chosen to ensure that the total population size remains close to $N$. In the mutation step, each individual mutates with probability $\Ud + \Un$, and if it does, the new mutation is deleterious with probability $\Ud/(\Ud+\Un)$. This process is continued for a sufficiently long period of time that the population reaches the steady-state mutation-selection balance introduced in the main text, and several population-wide coalescence events have occurred. A copy of our implementation in C is available upon request.

\end{appendix}
\end{document}